\documentclass[%
 aip,
 amsmath,amssymb,
reprint,%
]{revtex4-1}

\usepackage{graphicx}
\usepackage{dcolumn}
\usepackage{bm}
\usepackage[utf8]{inputenc}
\usepackage[T1]{fontenc}
\usepackage{mathptmx}
\usepackage{etoolbox}
\usepackage{float}
\usepackage[separate-uncertainty=true]{siunitx}
\usepackage{graphicx}
\usepackage{subcaption}
\usepackage{caption}
\usepackage{natbib}
\usepackage{ragged2e}
\usepackage[absolute,overlay]{textpos}
\usepackage{hyperref}
\hypersetup{
    colorlinks=true,
    linkcolor=blue,  
    citecolor=blue,  
    urlcolor=blue,   
}

\makeatletter
\def\@email#1#2{%
 \endgroup
 \patchcmd{\titleblock@produce}
  {\frontmatter@RRAPformat}
  {\frontmatter@RRAPformat{\produce@RRAP{*#1\href{mailto:#2}{#2}}}\frontmatter@RRAPformat}
  {}{}
}%
\makeatother
\begin{document}

\preprint{AION-REPORT/2023-10}

\title{Progress towards ultracold Sr for the AION project – sub-microkelvin atoms and an optical-heterodyne diagnostic tool for injection-locked laser diodes}
\author{E. Pasatembou}
\author{C. F. A.
  Baynham}%

\author{O. Buchm\"uller}
\altaffiliation[Also at ]{University of Oxford, Department of Physics, Clarendon Laboratory, South Parks Road, Oxford OX1 3PU, UK}
\author{D. Evans}

\author{R. Hobson}

\author{L. Iannizzotto-Venezze}

\author{A. Josset}

\email{c.baynham@imperial.ac.uk}
\affiliation{
  Department of Physics, Blackett Laboratory, Imperial College, Prince Consort Road, London, SW7 2AZ, U.K.
}%

\date{\today}

\begin{textblock*}{4cm}[0.5,0](\textwidth,1cm)
  \text{AION-REPORT/2023-10}
\end{textblock*}

\begin{abstract}
  Long-baseline atom interferometers, such as the one to be built by the AION collaboration, require ultra-cold atomic clouds. These are produced by trapping the atoms in Magneto-Optical Traps (MOTs) using high-power, narrow-linewidth lasers. We report on the laser and optical master-slave injection locked system used to address the $^1$S$_0$~--~$^3$P$_1$ strontium transition at \SI{689}{\nano\meter}, and on the trapping of strontium atoms in a narrowband MOT. We demonstrate the quality of the injection through the characterisation of the injection lock using a novel, easy-to-assemble method which uses a double pass acousto-optic modulator (AOM) to generate and detect a heterodyne beatnote. The reported system is used to produce an atomic cloud at a temperature of 812~$\pm$~\SI{43}{\nano\kelvin} in a narrowband red MOT.

\end{abstract}

\maketitle

\section{\label{sec:level1}introduction}

The Atom Interferometer Observatory and Network (AION) project \cite{Badurina2020} aims to use cold strontium atoms and atom interferometry to search for ultra-light dark matter and for gravitational waves (GWs) in the mid-frequency band ($\sim$ 0.01 Hz to a few Hz). Possible sources of these GWs include intermediate-mass black holes to which current and planned experiments including LISA \cite{lisa}, LIGO \cite{ligo} and Virgo \cite{virgo} are not sensitive. The NANOGrav collaboration has recently discovered evidence for low-frequency gravitational waves originating from supermassive black hole binaries providing further motivation for exploring different frequency bands \cite{Agazie2023}.

This staged project is to begin with a \qty{10}{\meter} baseline atom interferometer which will pave the way to a \qty{100}{\meter} detector and eventually a km-scale terrestrial detector. The final stage is the development of a satellite-based detector similar to the proposed AEDGE mission \cite{aedge}. The project is currently in Stage 1, with the construction of the \qty{10}{\meter} detector planned in Oxford.

\begin{figure}[h!]
  \includegraphics[width=1\columnwidth, height=0.45\textheight]{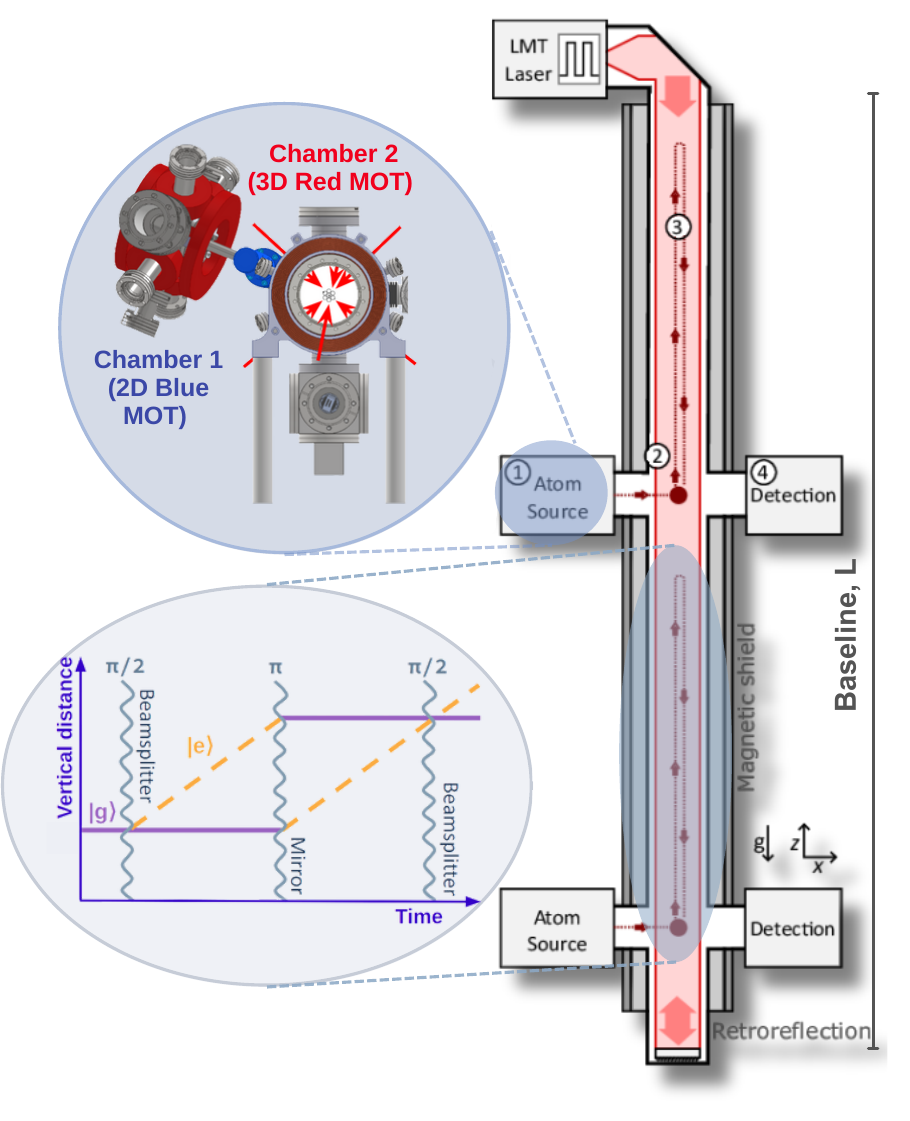}
  \caption{\label{fig:detector} \justifying
  Conceptual design of the AION detector. The atom cloud is cooled down in the atom source sections, transported into a long vertical tube, and then launched and allowed to free-fall in the tube where atom interferometry takes place. A sequence of pulses will split and recombine the two atomic paths. A differential phase measurement reveals information about the interaction of the atoms at the end of the interferometric sequence. Figure adapted from: [\onlinecite{Badurina2020}]}
\end{figure}

The AION detector will operate using single-photon strontium atom interferometry~\cite{Graham2016}. A schematic diagram of the detector is shown in Figure \ref{fig:detector}. In the \qty{10}{\meter} detector a common laser source will be used to operate two atom interferometers simultaneously. In subsequent stages, several atom interferometers will be used all addressed by a common laser source and large momentum transfer techniques will be used \cite{Rudolph2019}. A differential phase measurement between the interferometers will be taken at the end of the interferometric sequence \cite{Snadden1998}. To permit long flight times, sub-microkelvin temperatures are required for the atomic cloud before launch, as well as the mitigation of systematic effects~\cite{Dimopoulos2008}.

To reach the required temperature, a sequence of cooling techniques is used. This includes a 2D "blue" Magneto-Optical Trap (MOT), a 3D "blue" MOT, a 3D "red" MOT and an optical dipole trap for evaporative cooling \cite{Stellmer2013,Campbell2017}. The atoms are then optically transported to the interferometry region where they are launched and interferometry takes place. The set of Ultra-High Vacuum (UHV) chambers where the atoms are cooled down, or so-called sidearms, are attached to the side of the vertical detector next to the interferometer tube as shown in Figure \ref{fig:detector}. More details on the sidearm design and production are presented in Ref.~[\onlinecite{centralised}].

The AION project uses strontium, an alkaline-earth atom, due to its narrow-linewidth $^1$S$_0$~--~$^3$P$_0$ clock transition at \SI{698}{\nano\meter} used for interferometry \cite{Hu2017, Rudolph2019} and convenient optical transitions which allow for efficient trapping in MOTs \cite{Katori1999, Takashi2003, Xu2003, Sorrentino2006}. For the first stage of cooling, the 2D and the 3D ``blue'' MOT, the strong, dipole-allowed $^1$S$_0$~--~$^1$P$_1$ transition at \SI{461}{\nano\meter} is used. This transition has a natural linewidth of \SI{32}{\mega\hertz}. The subsequent cooling stage, the ``red MOT'', operates on the intercombination transition $^1$S$_0$~--~$^3$P$_1$ at \SI{689}{\nano\meter}. This is a narrow transition ($\gamma$ = \SI{7.6}{\kilo\hertz}) and can be used to reduce the temperature of the atomic sample down to half the photon recoil limit temperature of \SI{230}{\nano\kelvin} \cite{Katori1999, Loftus2004}.

The AION project requires the $^{87}$Sr isotope since its nuclear spin permits driving the \SI{698}{\nano\meter} clock transition through mixing of states via the hyperfine interaction.
However, the bosonic $^{88}$Sr isotope was used in this work since the isotope's natural abundance and lack of hyperfine structure make initial trapping easier.
The results and experimental details of the preceding 2D and 3D ``blue'' MOTs were presented in Ref.~[\onlinecite{centralised}]; this paper will focus on the second cooling stage, the red MOT.

This paper presents the first results from the narrowband 3D "red" MOT at Imperial College London in Section \ref{sec:results}. The optical setup used to achieve a red MOT is then presented and discussed in Section \ref{sec:lasersystem}. Section \ref{sec:injection} discusses how optical injection is achieved and introduces a novel way of characterising the injection lock using heterodyne beat detection.

\begin{figure}
  \includegraphics[width=\columnwidth]{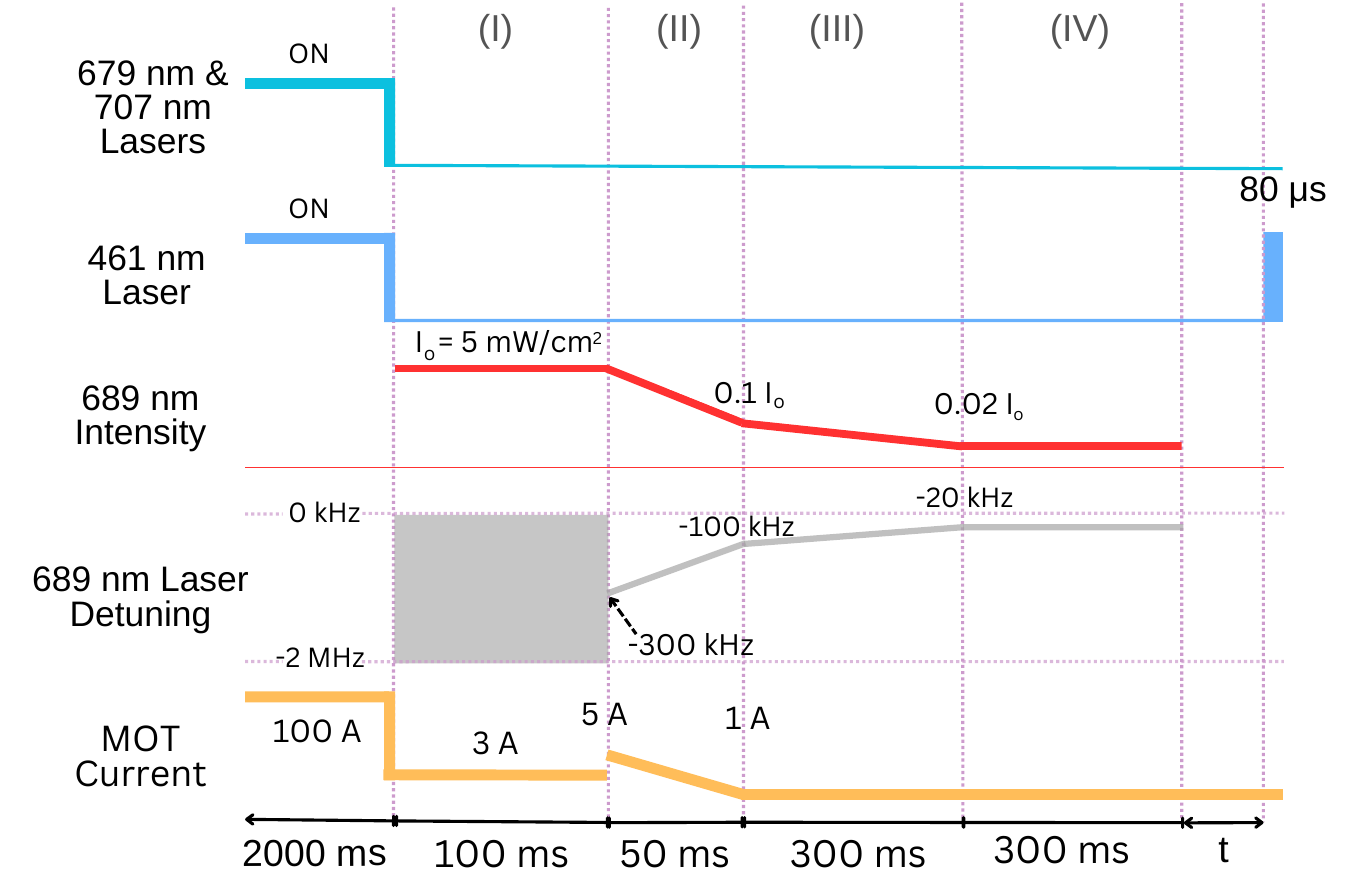}
  \caption{\label{fig:sequence} \justifying
  Cooling sequence timing diagram. The 3D blue MOT is loaded for \SI{2000}{ms} while the quadrupole coils are driven with a current of \SI{100}{A}, which gives a magnetic field gradient of approximately \SI{3.9}{\milli\tesla\per\centi\meter} (calibrated using measurements taken on identical coils in Ref.~[\onlinecite[\S3.2.3]{centralised}]).
  The \SI{461}{\nano\meter} laser and therefore the blue MOT is then turned off and the \SI{689}{\nano\meter} laser is turned on (with initial intensity $ I_{\text{o}}$ = \SI{5}{\milli\watt\per\centi\meter\squared}) transferring the atoms to the red broadband MOT over 100 ms (stage I). The current is reduced to 3 A over this step while the frequency is modulated to broaden the transition. In stage (II), the frequency modulation is turned off and atoms are trapped in a single-frequency MOT. The cloud is compressed by decreasing the detuning and intensity of the beam as well as the MOT current. The atoms are then held in the narrowband red MOT for 300 ms (stage (IV)). The cloud is led to expand for a variable time (t). The detection of Sr atoms for monitoring the dynamics of the red MOT is done using the \SI{461}{\nano\meter} transition. A probe beam interacts with the atoms and the resulting fluorescence is detected via CMOS cameras (BFS-PGE-50S5M-C PoE GigE Blackfly S Monochrome camera). The fluorescence pulse and the camera exposure are on for \SI{80}{\micro\second}. The setup includes two cameras, imaging the atom cloud in the horizontal x-y plane and the plane along gravity (x-z plane). }
\end{figure}

\section{\label{sec:results}Red MOT Sequence and Temperature Characterisation}

In this section, we present the results for the temperature reached in the narrowband red MOT. A brief description of the experimental sequence is also presented.

Figure \ref{fig:sequence} shows the laser and magnetic field cooling sequence used both to load and to characterise the temperature of the atom cloud trapped in a red narrowband MOT. As described in Ref.~[\onlinecite{centralised}], a 3D blue MOT is loaded by applying Doppler-cooling light at \SI{461}{\nano\meter} and repumping light at \SI{679}{\nano\meter} and \SI{707}{\nano\meter}, addressing the $^3$P$_0$~--~$^3$S$_1$ and $^3$P$_2$~--~$^3$S$_1$ transitions respectively.
For the initial broadband stage of the red MOT, we apply three circularly polarised, orthogonal, retro-reflected beams, each with a \SI{9}{\milli\meter} 1/e\textsuperscript{2} radius and \SI{5}{\milli\watt\per\centi\meter\squared} intensity. The beams retroreflect through an achromatic quarter-waveplate suitable for both blue and red wavelengths. One beam propagates coaxially with the paired anti-Helmholtz coils that create the MOT field gradient~[\onlinecite[\S3.2.3]{centralised}]. The other two propagate at \SI{45}{\degree} angles to gravity.

To increase the efficiency of atom transfer from the blue to red MOT cooling stages, we apply saw-tooth modulation to the red light's frequency to effectively broaden the narrow transition and address a wider range of velocity classes in the atomic cloud. We apply a positive ramp from detuning $\delta=-\SI{2}{\mega\hertz}$ to $\delta=\SI{0}{\mega\hertz}$, repeated at \SI{30}{\kilo\hertz}.

After initial capture in the broadband MOT, we transition to ``capture'' and ``compression'' phases in which the \SI{689}{\nano\meter} light is no longer frequency modulated and the beam detuning, beam intensities and field gradient are linearly ramped as shown in Figure~\ref{fig:sequence}. These ``narrowband'' stages result in a cooler and more compressed cloud (sub-millimeter scale) suitable for later loading into a crossed optical dipole trap. We note that the sequence parameters shown in Figure~\ref{fig:sequence} have not been exhaustively optimised; with more thorough searches of the sequence parameter space we anticipate improvements in both atom number and ultimate temperature. 

\begin{figure}[tb]
  \centering
  \includegraphics[width=1\columnwidth]{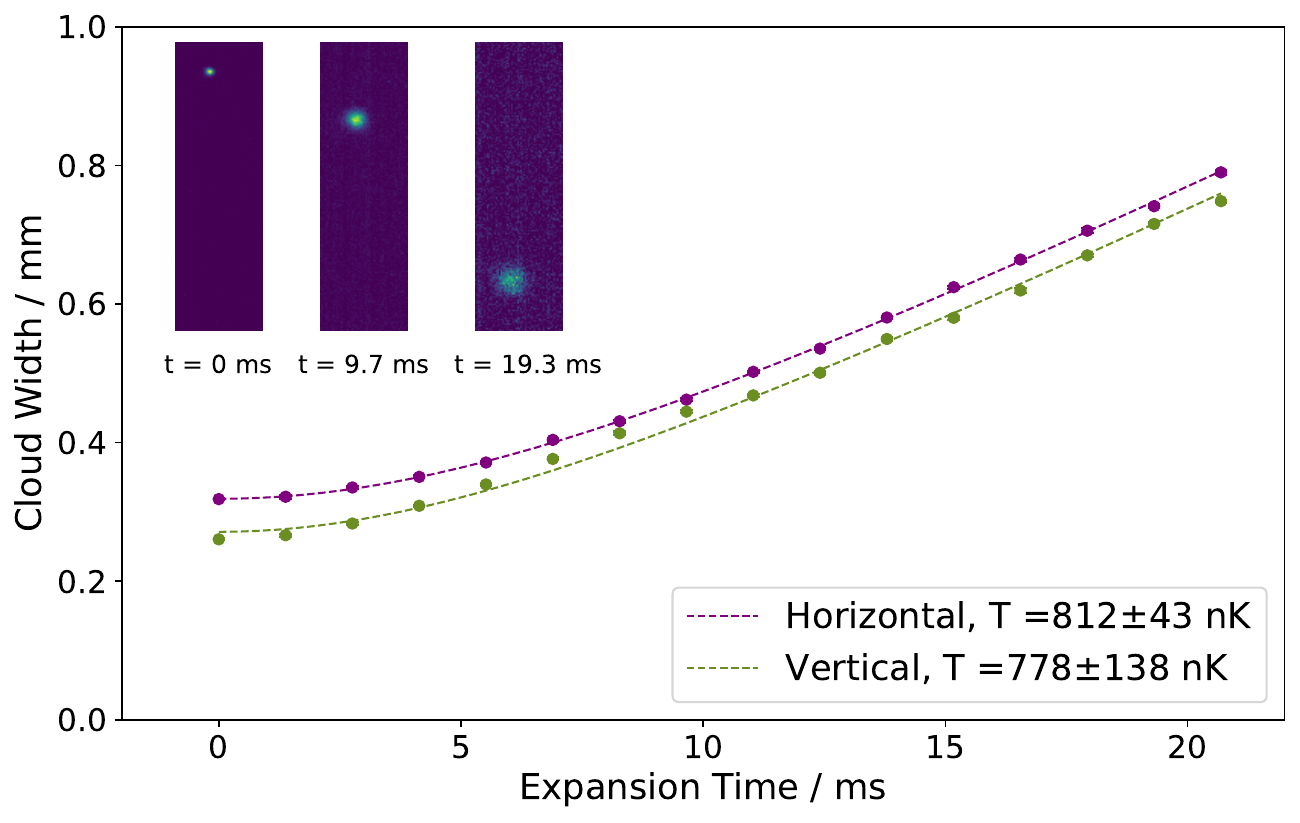}
  \caption{
    \justifying
    Time-of-flight measurements of the atomic cloud taken after the atoms are released from the final red MOT stage. Each datapoint consists of 30 randomised repeats whose standard error is taken as the statistical uncertainty. 
    \textit{(inset)} Sample raw images of the expanding atomic cloud.
  }
  \label{fig:expansion}
\end{figure}

Time-of-flight (TOF) fluorescence imaging is then used to determine the temperature of the atomic cloud. We image in a plane parallel to gravity so that the known acceleration of the cloud under free-fall can be used to calibrate the size of one pixel in the image plane (see Figure~\ref{fig:expansion}). Using this technique, we determine the camera resolution to be \SI{30.9(1)}{\micro\meter}/pixel. 
For the TOF images, the \SI{689}{\nano\meter} light is turned off then the atoms are allowed to freely expand for variable amounts of time $t$ between \SIrange{0}{20.7}{\milli\second}. Images are taken by pulsing the \SI{461}{\nano\meter} MOT beams for \SI{80}{\micro\second} synchronously with a gating trigger pulse to our camera, optically filtered by an FBH460-10 band-pass filter centred around \SI{460}{\nano\meter}. 

A 2D Gaussian fit was applied to the TOF images to determine cloud widths at variable expansion times. The relationship between the temperature $T$ and the $1/e^2$ radius $r(t)$ of the cloud is given by~\cite{Brzozowski2002}
\begin{equation}
  \label{eq:temperature}
  T = \frac{m}{4k_Bt^2}\left[r(t)^2 - r(0)^2\right]
  \quad,
\end{equation}
\noindent where $m$ is the mass of a strontium atom and $k_B$ is Boltzmann's constant. $r(0)$ is the radius of the cloud at time $t = 0$ before the cloud starts expanding and, with $T$, is a free parameter of the fit.

Figure~\ref{fig:expansion} shows an example of three TOF measurements, as well as the result of fitting Equation~\ref{eq:temperature} to the measured cloud widths in the horizontal and vertical (i.e. along gravity) directions. We find horizontal and vertical temperatures in the final narrowband red MOT stage of \SI{812(43)}{\nano\kelvin} and \SI{778(138)}{\nano\kelvin} respectively. The $1/e^2$ widths of the unexpanded cloud are \SI{319(1)}{\micro\meter} and \SI{261(1)}{\micro\meter} in the horizontal and vertical directions respectively.

\begin{figure*}
  \includegraphics[width=0.95\textwidth, height=0.36\textheight]{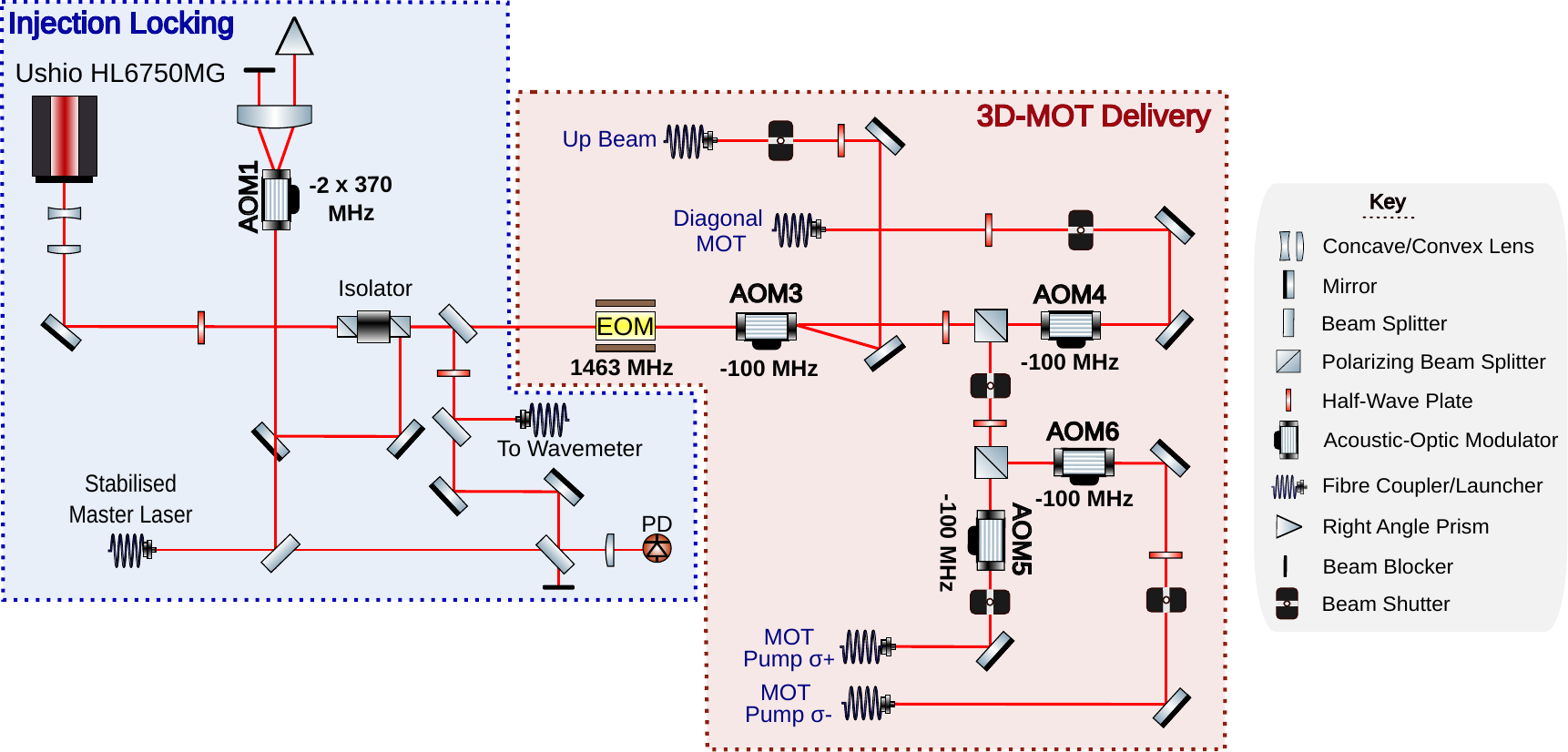}
  \caption{\label{fig:689} Laser system and optical setup for the delivery of light to the 3D red MOT.}
\end{figure*}

\section{\label{sec:lasersystem}Laser system}

In this section, we describe the laser system used to obtain the optical beams required to achieve a narrow-band red MOT. To reach temperatures below $\sim$ \SI{1}{\micro\kelvin}, the red MOT requires \SI{689}{\nano\meter} laser light with $\lesssim 1$~kHz linewidth. The light is created and delivered to the 3D MOT chamber using the optical setup shown in Figure \ref{fig:689}, which includes two main parts: injection locking and the 3D MOT laser light delivery optics. The \SI{689}{\nano\meter} master laser light is supplied to the laser stabilisation and MOT delivery systems by a \SI{689}{\nano\meter} Toptica DL Pro ECDL, which has a maximum power output of around \SI{25}{\milli\watt} and a specified free-running linewidth of \SI{50}{\kilo\hertz} at \SI{5}{\micro\second}. The master ECDL is frequency stabilised to an ultra-low expansion glass optical cavity and its linewidth is narrowed via the Pound-Drever-Hall technique \cite{Black2001, centralised}.

For optimal operation of the frequency-modulated broadband red MOT, the intensity of the red light should be several hundred times the transition's saturation intensity (\SI{3}{\micro\watt\per\centi\meter\squared}) - more than can be delivered from the master laser alone.~\cite{Bartolotta2018, Snigirev2019}
To achieve this, a Ushio HL6750MG diode is included in the delivery system, which has a maximum power output of around \SI{50}{\milli\watt} and a nominal central lasing wavelength of \SI{685}{\nano\meter}. This diode is injection-locked to the master light with a controllable offset. 

The diode is housed in a temperature-stabilised aluminium enclosure. Light from the cavity-stabilised master laser is passed through a double-pass acousto-optic modulator (AOM) system and an optical isolator and then injects the slave laser diode. AOM1 permits fast control of the MOT beam frequency, used for example in the broadband MOT stage. Cylindrical lenses are used to shape the laser diode output into a circular beam. Once the injected beam passes through the isolator, it proceeds to take two paths, one through the rest of the MOT delivery system and the other to the injection characterisation setup. Details on injection locking and how it is achieved and characterised are discussed in Section \ref{injection}.

The 100 MHz AOMs are used for active intensity control and are driven by a SUServo module from the ARTIQ experimental control system (an FPGA-based control system for quantum information experiments~\cite{artiq}). This provides FPGA-based PI control of the beam intensity delivered to the chamber.

For the purpose of this paper, $^{88}$Sr is used for the initial testing of the MOTs. In future experiments and in AION, $^{87}$Sr will be used which adds additional requirements for the laser system to address the hyperfine structure levels. To meet these requirements, the frequency of the \SI{689}{\nano\meter} light will be modulated by a resonant free-space electro-optic modulator (EOM) driven at \SI{1463}{\mega\hertz}. The +1 frequency sideband of the EOM will be tuned to address the $F = 9/2 \rightarrow 9/2$ transition, resulting in “stirring” of the red MOT, a method used to redistribute the population in the m$_F$ sublevels and increase the efficiency of trapping \cite{Mukaiyama2003}. The carrier beam will be used to target the $F = 9/2 \rightarrow 11/2$ transition for trapping.

\section{\label{sec:injection}Injection Lock Characterisation}  \label{injection}

Injection locking refers to the technique whereby a laser is forced to operate on a specific frequency by injecting light of the required frequency into the laser resonator \cite{Stover1966, Yuan2020, Liu2020}. The low-noise, low-power Toptica laser is injected into the high-power slave laser to get a high-power and low-noise resulting laser beam of the required frequency and narrow linewidth \cite{Fleming1981}. For the diode to be injected it needs to be within certain cavity mode bounds to get a stable injection lock. There is a locking range defined by the injection field amplitude related to the peak diode cavity mode and the mode spacing. In this case, the peak cavity mode can be tuned by changing the current supplied to the diode by the diode controller or by changing the diode's temperature.

\begin{figure}
  \includegraphics[width=\columnwidth]{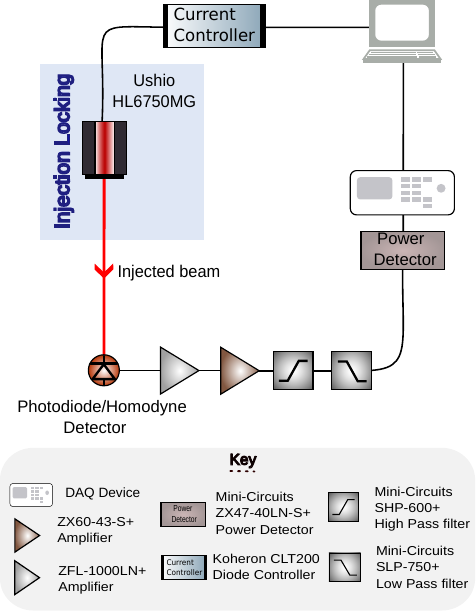}%
  \caption{\label{fig:injection}
  \justifying
  Setup for achieving slave injection and characterisation of the injection lock. The optical system used to achieve injection and get a beat note is shown in the blue box. The rest of the system is used to control the diode current and temperature and perform current sweeps for the characterisation of the injection lock.}
\end{figure}

In this section, we present a novel method used to achieve and characterise the injection lock in our laser system. This method is simple and easy to assemble as it uses an AOM to produce a heterodyne beat between the injected and master laser beams. The beat is then used characterise the injection. Figure \ref{fig:injection} shows the setup used for laser injection and characterisation of the injection lock.

The master laser and the injected slave laser beam are coupled to a high-resolution HighFinesse WS8-10 wavemeter for frequency monitoring. The position of the master beam is aligned geometrically to overlap with the counterpropagating slave beam, and the diode current is then tuned until injection is achieved, which is monitored by observing the frequency of the two beams on the wavemeter. The frequency of the injected beam is 680 MHz (frequency change due to the double pass AOM) less than that of the master laser frequency.

\begin{figure}
  \begin{subfigure}{1\columnwidth}
    \includegraphics[width=1\linewidth]{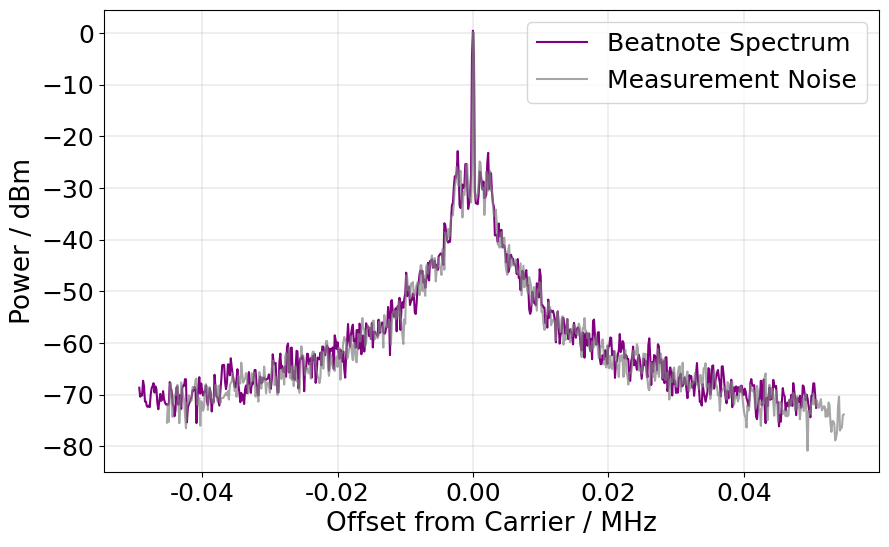}
    \caption{}
    \label{fig:beatnote}
  \end{subfigure}

  \begin{subfigure}{1\columnwidth}
    \includegraphics[width=1\linewidth]{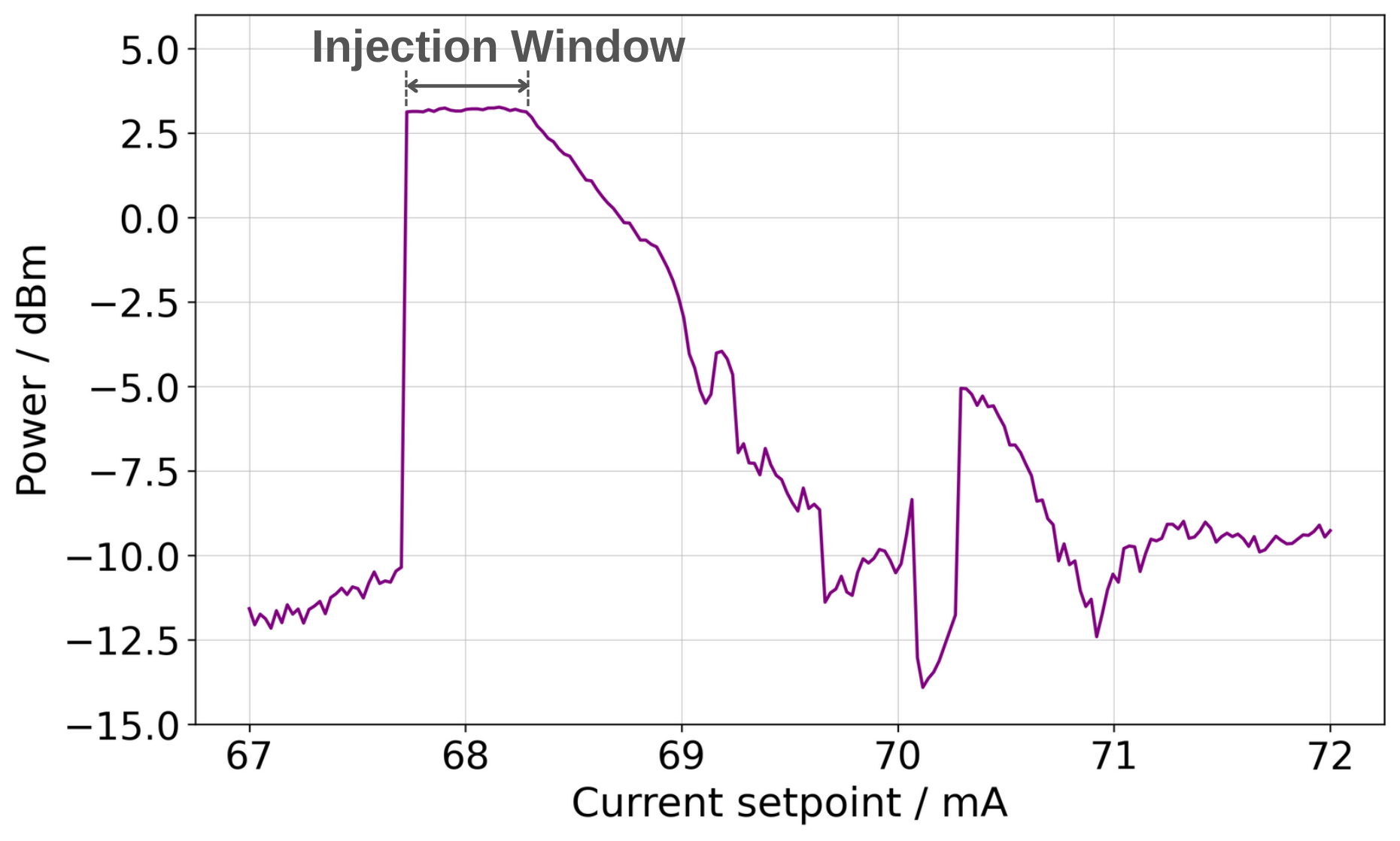}
    \caption{}
    \label{fig:window}
  \end{subfigure}

  \caption{%
  \justifying
  \textit{(a)} The beat spectrum of the heterodyne signal measured by the photodiode using a bandwidth of \SI{100}{\kilo\hertz}. The beatnote spectrum is centered around the carrier frequency which is the difference in frequency between the master and the injected beams. A background noise trace was also taken of an independent, high stability \SI{680}{MHz} RF source revealing a source of measurement noise. The authors suspect phase noise within the spectrum analyser but note that the injected signal is indistinguishable from the ideal case within the precision of our instrument. 
  \textit{(b)} Injection characterization data as taken from the power detector.
  The injection signal exhibits strong hysteresis, as is common with injected systems.
  This scan was performed downwards in current over the setpoint ($\sim$ \SI{68.0}{\milli\ampere}). The flat region with the maximum power is the range of currents over which the diode is injected. This injection window is around \SI{0.6}{\milli\ampere} wide. This scanning trace was used to optically align the system and mode-match the injection beam to the diode's output by maximising the width of the injection window as changes were made.}
  \label{fig:merged}
\end{figure}

To diagnose and optimise the injection more carefully, a photodiode is used to detect the beat note created by combining the master laser with the injected beam. The beat note can be displayed on a spectrum analyser for a first evaluation of the injection. An example of a beat note showing injection is shown in Figure \ref{fig:beatnote}.

The injection lock can be characterised using the configuration shown in Figure \ref{fig:injection}. This configuration allows the heterodyne beat signal to be monitored on an oscilloscope and for the voltage of the power detector input signal to be measured as a function of time. A triangle waveform is generated which is used to sweep the current of the laser diode and produce a plot of the input power as a function of current as shown in Figure \ref{fig:window}. The range of currents over which the injection stays locked can therefore be found.

The diode was operated using the maximum possible current (around \SI{70}{\milli\ampere}), producing a \SI{50}{\milli\watt} beam. The slave diode stays injected over a window of $\sim$ \SI{0.6}{\milli\ampere} width (from \SIrange{68.3}{67.7}{\milli\ampere}).

This method has many benefits over other methods found in the literature, e.g., a scanning cavity. It can easily be assembled using an AOM, a photodiode and a power detector as the main components to detect the heterodyne beat. The method introduced also allows for the measurement of the injection lock window which is achieved by scanning the slave diode laser current and observing the range of currents over which the injection stays locked. This allows for the assessment of the quality of the injection, and can be used to carefully optimise beam alignment and diode temperature to maximise the injection window, and therefore the robustness of the injection lock.

\section{Conclusion}  \label{conclusion}

This paper presents the first results of a narrowband red MOT operated on $^{88}$Sr atoms at Imperial College London. An atomic cloud was cooled down to a temperature of \SI{812(43)}{\nano\kelvin} in a narrowband red MOT. This temperature is low enough to efficiently load an optical dipole trap - the next step towards differential atom interferometry. This is an important milestone for the AION collaboration as we move from tabletop experiments to the realisation of a long baseline detector used to detect gravitational waves, ultra-light dark matter and beyond.

We have also presented the laser and optical setup used to realise the narrowband red MOT. The laser system is a simple and easy-to-assemble master-slave injection-locked system, capable of generating $\sim$ \SI{50}{\milli\watt} at \SI{689}{\nano\meter}. We have demonstrated a novel way of characterising the injection lock by using an AOM to generate a beatnote by combining the master and injected beam. The injection lock system was characterised using a beat note spectrum and by performing a current sweep and determining an injection lock window showing the range over which the slave laser stays injected. We have demonstrated that the slave diode stays injected over a range of \SI{0.6}{\milli\ampere}.

\begin{acknowledgments}

  This research is funded by UKRI as part of the Quantum Technology for Fundamental Physics program, through grants provided by EPSRC and STFC within the AION Consortium framework. The specific grant references are ST/T006994/1 and ST/W006332/1.

  The authors thank Dan Reed for the original diode enclosure design, and Tiffany Harte and Maurice Zeuner for adapting the design for the AION project injected diodes, as well as the entire AION collaboration for support and collaboration throughout all stages of the experiment. 

  The authors are also grateful to David Nadlinger and the Oxford Ion Trapping Group for the excellent \href{https://github.com/OxfordIonTrapGroup/ndscan}{ndscan package} (\url{https://github.com/OxfordIonTrapGroup/ndscan}).

\end{acknowledgments}

\section*{Author Declarations}
\subsection*{Conflict of Interest}
The authors have no conflict to disclose.

\section*{Data Availability}

The data that support the findings of this study are available from the corresponding author upon reasonable request.

\section*{References}
\nocite{*}
\bibliography{main}

\end{document}